\documentclass[10pt,aps,showpacs,twocolumn]{revtex4}
\usepackage{amsmath}
\usepackage{bm}
\usepackage{epsfig}
\def\be{\begin{equation}}
\def\ee{\end{equation}}
\def\bea{\begin{eqnarray}}
\def\eea{\end{eqnarray}}

\begin{document}

\title{{\large Effects of Chirped Laser Pulses on Nonclassical Correlation and
Entanglement of Photon Pairs from Single Atom }}
\author{C. H. Raymond Ooi{\footnote{\sl e-mail address:bokooi73@yahoo.com}}}
\affiliation{\sl Department of Physics, Korea University,
Anam-dong, Seongbuk-gu, Seoul, 136-713 Republic of Korea}
\date{\today}

\begin{abstract}
We study the effects of arbitrary laser pulse excitations on
quantum correlation, entanglement and the role of quantum noise.
The transient quantities are computed exactly using a method that
provides exact solutions of the Langevin field operators for
photon pairs produced by a double Raman atom driven by laser
pulses. Short pulses with appropriate chirping, delay and/or
detuning can generate broadband photon pairs and yield results
that provide insights on how to generate very large nonclassical
correlation. We find that short pulses are not favorable for
entanglement. The quantity was previously found to be
phase-sensitive and this is used with the pulse area concept to
explain the rapid variations of entanglement with pulse width and
strength. Photon correlation and entanglement are favored by
exclusively two different initial conditions. Analysis reinforces
our understanding of the two nonclassical concepts.
\end{abstract}

\maketitle

\section{Introduction}
The intriguing nature of atoms and photons manifest through
entanglement and quantum correlation. These nonclassical concepts
carry subtly different physical meanings although they are
mathematically related for two-mode Gaussian state
\cite{entanglement correlation}. As pointed out by R. J. Glauber,
quantum correlation exists when the density matrix is not a
product form $\prod_{k}|\alpha _{k}\rangle \langle \alpha _{k}|$
(in coherent state basis), rather in an incoherent superposition
$\sum_{k}w_{k}|\alpha _{k}\rangle \langle \alpha _{k}|$ or mixed
state \cite{Glauber}. Entanglement is commonly depicted as a
superposition of two mutually inverted states. The physical
mechanisms that create quantum correlation and entanglement may
not be the same and not fully understood.

\begin{figure}[tbp]
\center\epsfxsize=8cm\epsffile{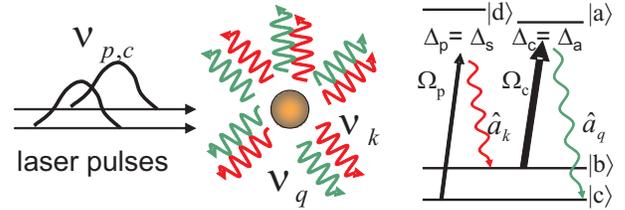} \caption{(Color
online) An atom with four-level double Raman scheme is
driven by pump(p) and control(c) laser pulses, producing Stokes ($\mathbf{k}$%
) and anti-Stokes ($\mathbf{q}$) photons with transient quantum
correlation and entanglement.} \label{pulsedscheme}
\end{figure}

We consider a single atom with four-level double $\Lambda $ system (Fig. \ref
{pulsedscheme}) driven by laser \emph{pulses}. This scheme is interesting
for being able to generate controllable nonclassically correlated Stokes and
anti-Stokes photons from an atom \cite{HH} as well as extended medium \cite
{paper I}. Previous works on single atom driven by continuous wave (c.w.)
lasers were based on Schr\"{o}dinger's equation \cite{HH}, \cite{Ooi} and
master equation \cite{Patnaik}. The former approach yields solutions only
for c.w. lasers. The latter uses atomic correlation to obtain field
correlations indirectly via quantum regression theorem \cite{qtm regr}. A
general and more transparent method for computing correlations is based on
Heisenberg-Langevin (HL) formulation, which can be extended to include
spatial propagation. However, the previous studies have not considered
excitations with laser pulses.

Laser pulses provide extra degrees of freedom for coherent control
of nonclassicality. It is interesting to explore the effects of
pulse width (duration) and other parameters such as chirp, phase
and shape on quantum correlation and entanglement of photon pairs,
as well as on the role of quantum noise \cite{role of noise}.
Nonclassical photon pairs driven by laser pulses could be made
sufficiently intense and serves as a new tool for doing
\emph{quantum nonlinear optics}. A scheme to generate single-cycle
photon pairs has been proposed \cite{Harris single cycle}. For
(short and intense) laser pulses, exact solutions cannot be
obtained using the Schr\"{o}dinger's equation. The quantum
regression approach may give numerical solutions for pulses, but
cannot account for spatial propagation. The ultimate goal is to
compute quantum correlation for extended medium driven by
arbitrary laser pulses.

As the first step towards this goal, we present a general method that gives
exact transient solutions for quantum operators ($\hat{a}_{k}$ and $\hat{a}%
_{q}$) of photon pairs in HL equations for single atom. The solutions enable
us to study, for the first time, the effects of various laser pulse
parameters on the transient quantum correlation and entanglement of the
photon pairs. We obtain interesting results that reflect the unique
advantages of using pulses, thus showing the inherent advances in extending
the previous studies to pulsed excitations. We find that chirped and
overlapping laser pulses can give an extremely large photon correlation. In
addition, short pulses tend to increase the photon correlation but reduces
the entanglement. We give physical explanations that shed insights on the
effects of finite spectral bandwidth on the nonclassical properties. By
analyzing the main results that contain rich underlying physics, we acquire
conceptual insights on: a) coherent control of nonclassical photons, b) the
differences between nonclassical correlation and entanglement of photon
pairs and b) the effects of pulses on the role of quantum noise.

\section{Theoretical Formulation}

The usual Hamiltonian for an atom in Fig. \ref{pulsedscheme} interacting
with radiation fields and two laser fields is

\begin{eqnarray}
\hat{V}_{I} &=&-\hbar \lbrack \Omega _{p}\hat{\sigma}_{ac}e^{-i\Delta
_{p}t}+\sum_{k}g_{k}^{\ast }\hat{a}_{k}\hat{\sigma}_{ab}e^{-i\Delta _{k}t}
\notag \\
&&+\Omega _{c}\hat{\sigma}_{ab}e^{-i\Delta _{c}t}+\sum_{q}g_{q}^{\ast }\hat{a%
}_{q}\hat{\sigma}_{ac}e^{-i\Delta _{q}t}+\text{adj.}]  \label{V}
\end{eqnarray}
where $\Omega _{p(c)}$ is the Rabi frequency of the pump (control) field, $%
\Delta _{k}=\nu _{k}-\omega _{ab}$, $\Delta _{q}=\nu _{q}-\omega _{ac}$ are
the detunings of the Stokes (subscript-k) and anti-Stokes (subscript-q)
frequencies, $g_{k}^{\ast }$, $g_{q}^{\ast }$ are the usual atom-field
coupling coefficients.

It is straightforward to obtain a set of sixteen HL equations including
noise operators as represented symbolically by

\begin{equation}
\frac{d}{dt}\tilde{\sigma}_{xy}(t)=-\Gamma _{xy}\tilde{\sigma}%
_{xy}(t)+i\sum\limits_{x,y}C_{mn}\tilde{\sigma}_{mn}+\tilde{F}_{xy}(t)
\label{HL}
\end{equation}
where $\tilde{F}_{xy}=\hat{F}_{xy}e^{-i\Delta _{xy}t}$ and $\Gamma
_{xy}=\gamma _{xy}+i\Delta _{xy}$ stands for the complex depopulation or
decoherence rates, $C_{mn}$ is the coefficient that depends on the laser
fields and $\hat{F}_{xy}$ represents the noise operators. The HL equations
represented by Eq. (\ref{HL}) are obtained after eliminating the solutions
of the field operators $\hat{a}_{k}$ and $\hat{a}_{q}$ in the Heisenberg
equation $\frac{d}{dt}\hat{O}=\frac{1}{i\hbar }[\hat{O},\hat{H}]$ for atomic
operators using
\begin{equation}
\frac{d}{dt}\tilde{\sigma}_{k}^{\dagger }(t)=-ig_{k}^{\ast }\tilde{\sigma}%
_{db}(t)  \label{dak/dt}
\end{equation}
\begin{equation}
\frac{d}{dt}\tilde{\sigma}_{q}(t)=ig_{q}\tilde{\sigma}_{ca}(t)
\label{daq/dt}
\end{equation}
where $\tilde{\sigma}_{k}(t)=\hat{a}_{k}e^{-i\nu _{k}t}$ and $\tilde{\sigma}%
_{q}(t)=\hat{a}_{q}e^{-i\nu _{q}t}$.

\section{Method for Exact Solutions}

Exact analytical solutions for $\hat{a}_{k}$ and $\hat{a}_{q}$ can be found
from the exact solutions of $\hat{\sigma}_{ab}(t)$ and $\hat{\sigma}_{ca}(t)$%
(since these equations are linear) but the solutions are very cumbersome and
not helpful. Adiabatic approximation may simplify the solutions but it will
considerably limit the range of operating parameters. We outline a method
which provides exact solutions for the field operators.

First, note that the time evolution of the density matrix elements can be
solved numerically as $X(t)=\langle \hat{X}(t)\rangle =U(t)\langle \hat{X}%
(0)\rangle $, where $\hat{X}(0)=\{\hat{\sigma}_{\alpha \beta }(0)\}$ where $%
\alpha \beta =aa,...ad,ba,...,bd,.....dd$ is the initial vector of the
atomic operators and $U(t)$ is the time-evolution matrix whose exact form
can only be obtained numerically from the solutions $\rho _{yx}=\langle \hat{%
\sigma}_{xy}\rangle $ of the density matrix equations. We then find the
exact solutions
\begin{eqnarray}
\hat{a}_{k}^{\dagger }(t) &=&\hat{a}_{k}^{\dagger }(0)-ig_{k}^{\ast
}\sum_{m=1}^{16}  \notag \\
&&\left[ K_{m}(t)\hat{X}_{m}(0)+\int_{0}^{t}K_{m}(t^{\prime })\hat{F}%
_{m}(t-t^{\prime })dt^{\prime }\right]  \label{ak sol}
\end{eqnarray}
\begin{eqnarray}
\hat{a}_{q}(t) &=&\hat{a}_{q}(0)+ig_{q}\sum_{n=1}^{16}  \notag \\
&&\left[ Q_{n}(t)\hat{X}_{n}(0)+\int_{0}^{T}Q_{n}(t^{\prime })\hat{F}%
_{n}(t-t^{\prime })dt^{\prime }\right]  \label{aq sol}
\end{eqnarray}
where $K_{m}(t)=\int_{0}^{t}U_{14,m}(t^{\prime })dt^{\prime }$ and $%
Q_{n}(t)=\int_{0}^{t}U_{9,n}(t^{\prime })dt^{\prime }$. This technique for
obtaining exact operator solutions is applicable to an arbitrary time
dependent laser fields interacting with an atom. It may be extended to
molecules in future works.

The single operator expectation values in thermal vacuum are $\langle \hat{a}%
_{k}^{\dagger }(t)\rangle =-ig_{k}^{\ast }\sum_{m=1}^{16}K_{m}(t)\langle
\hat{X}_{m}(0)\rangle $ and $\langle \hat{a}_{q}(t)\rangle
=ig_{q}\sum_{m=1}^{16}Q_{m}(t)\langle \hat{X}_{m}(0)\rangle $. From the
solutions Eqs. (\ref{ak sol}) and (\ref{aq sol}), we can compute the
expectation values for the products of two operators. For example,

\begin{equation*}
\left\langle \hat{a}_{q}(t)\hat{a}_{k}(t)\right\rangle
=-g_{k}g_{q}\sum_{m,n=1}^{16}[Q_{m}(t)K_{n}^{\ast }(t)\langle \hat{X}_{m}(0)%
\hat{X}_{n}^{\dagger }(0)\rangle
\end{equation*}
\begin{equation}
+\int_{0}^{t}Q_{m}(T-t^{\prime })K_{n}(t-t^{\prime })2D_{mn}^{an}(t^{\prime
})dt^{\prime }]  \label{aqak sol}
\end{equation}
\begin{equation*}
\bar{n}_{k}=\bar{n}_{k}^{th}+|g_{k}|^{2}\sum_{m,n=1}^{16}[K_{m}(t)K_{n}^{%
\ast }(t)\langle \hat{X}_{m}(0)\hat{X}_{n}^{\dagger }(0)\rangle
\end{equation*}
\begin{equation}
+\int_{0}^{t}K_{m}(t-t^{\prime })K_{n}^{\ast }(t-t^{\prime
})2D_{mn}^{an}(t^{\prime })dt^{\prime }]  \label{nk sol}
\end{equation}
where $\bar{n}_{k}=\langle \hat{a}_{k}^{\dagger }(t)\hat{a}_{k}(t)\rangle $
is the Stokes photon number, $\langle \hat{a}_{k}^{\dagger }(0)\hat{a}%
_{k}(0)\rangle =\bar{n}_{k}^{th}$ and $\langle \hat{a}_{q}^{\dagger }(0)\hat{%
a}_{q}(0)\rangle =\bar{n}_{q}^{th}$ are the mean thermal photon numbers at
initial time. The normal ordered and antinormal ordered diffusion
coefficients are defined as $D_{mn}^{n}(t)=$ $\langle \hat{F}_{m}^{\dagger
}(t)\hat{F}_{n}(t)\rangle $ and $D_{mn}^{an}(t)=$ $\langle \hat{F}_{m}(t)%
\hat{F}_{n}^{\dagger }(t)\rangle $ respectively. The final terms that
contain the diffusion coefficients are the noise (superscript ``n'') parts,
i.e. $\bar{n}_{k}^{n}$ and $\bar{n}_{q}^{n}$, while the former terms are
referred as the boundary parts $\bar{n}_{k}^{b}$ and $\bar{n}_{q}^{b}$. We
also use $\hat{\sigma}_{\alpha \beta }\hat{\sigma}_{\beta \gamma }=(\hat{%
\sigma}_{\gamma \beta }\hat{\sigma}_{\beta \alpha })^{\dagger }=\hat{\sigma}%
_{\alpha \gamma }$. We have assumed that initially the modes are
statistically independent: $\langle \hat{a}_{q}(0)\hat{a}_{k}(0)\rangle
=\langle \hat{a}_{q}(0)\hat{a}_{k}^{\dagger }(0)\rangle =0$ and are not
squeezed, i.e. $\langle \hat{a}_{k}^{2}(0)\rangle =\langle \hat{a}%
_{q}(0)\rangle =0$.

\section{Nonclassical Quantities}

The average value of the paired operators are computed to obtain mean photon
numbers, photon-photon correlation and to determine entanglement. The
effects of finite pulse duration, chirping and pulse sequence on these
quantities can be studied. We also analyze the effects of pulses on the
contributions of the noise operators compared to the initial operators.

\subsection{Nonclassical Photon Correlation}

The solutions Eqs. (\ref{ak sol}) and (\ref{aq sol}) are linear combinations
of initial field operators and atomic noise operators. Thus, the two-photon
correlation function can be decorrelated into terms with paired operators.
The presence of nonclassical two-photon correlation between the Stokes and
anti-Stokes photons is determined by the condition $g^{CS}>1$ for the
Cauchy-Schwarz correlation
\begin{eqnarray}
g^{CS}(t) &=&\frac{\left\langle \hat{a}_{k}^{\dagger }(t)\hat{a}%
_{q}^{\dagger }(t)\hat{a}_{q}(t)\hat{a}_{k}(t)\right\rangle }{\sqrt{%
G_{k}^{(2)}(t)G_{q}^{(2)}(t)}}  \notag \\
&=&\frac{|\left\langle \hat{a}_{q}\hat{a}_{k}\right\rangle
|^{2}+|\left\langle \hat{a}_{q}^{\dagger }\hat{a}_{k}\right\rangle
|^{2}+\left\langle \hat{n}_{k}\right\rangle \left\langle \hat{n}%
_{q}\right\rangle }{\sqrt{\{|\langle \hat{a}_{k}^{2}\rangle
|^{2}+2\left\langle \hat{n}_{k}\right\rangle ^{2}\}\{|\langle \hat{a}%
_{q}^{2}\rangle |^{2}+2\left\langle \hat{n}_{q}\right\rangle ^{2}\}}}
\label{gCS}
\end{eqnarray}
where $G_{j}^{(2)}(t)=\left\langle \hat{a}_{j}^{\dagger }(t)\hat{a}%
_{j}^{\dagger }(t)\hat{a}_{j}(t)\hat{a}_{j}(t)\right\rangle $ ($j=k,q$) and
we omit the $t$ dependence to simplify the notations. Note that Eq. (\ref
{gCS}) corresponds to transient coincident (joint) two-photon detection
(with no delay between the Stokes and anti-Stokes) which is mathematically
related to an entanglement criteria Eq. (\ref{relate}) below.

\subsection{Entanglement}

The above solutions Eqs. (\ref{ak sol}) and (\ref{aq sol}) are also used to
identify the presence of entanglement via $D(t)=\langle {\left( {\Delta \hat{%
u}}\right) ^{2}\rangle +\langle \left( {\Delta \hat{v}}\right) ^{2}\rangle }%
<2$ \cite{Duan}. This is a sufficient and necessary to criteria for two-mode
Gaussian states. We compute the paired operators: $\left\langle \hat{a}%
_{q}(t)\hat{a}_{k}(t)\right\rangle $, $\langle \hat{a}_{k}^{\dagger }(t)\hat{%
a}_{k}(t)\rangle $, $\langle \hat{a}_{q}^{\dagger }(t)\hat{a}_{q}(t)\rangle $%
, $\langle \hat{a}_{q}(t)\hat{a}_{k}^{\dagger }(t)\rangle $, $\langle \hat{a}%
_{k}^{2}(t)\rangle $, $\langle \hat{a}_{q}^{2}(t)\rangle $ just like Eqs. (%
\ref{aqak sol}) and (\ref{nk sol}). Note that the finite mean thermal photon
numbers $\bar{n}_{k}^{the}$ or $\bar{n}_{q}^{the}$ effectively increases the
value $D$. Thus, entanglement may not be obtained when there is a
significant number of thermal photons.

Let use recall the steady state for two-photon state $\left| \Phi
\right\rangle =\sum_{\mathbf{k},\mathbf{q}}C_{\mathbf{k}\mathbf{q}}\left|
c,1_{\mathbf{k}},1_{\mathbf{q}}\right\rangle $ \cite{HH}. For Raman-EIT
scheme which shows nonclassical correlation, the coefficient is \emph{%
inseparable }$C_{\mathbf{k}\mathbf{q}}\neq C_{\mathbf{k}}C_{\mathbf{q}}$
\cite{Ooi}. Thus, one might expect to find entanglement in the steady state
as well as in the transient. However, we find no entanglement for the
Raman-EIT scheme at all times and with any control field $\Omega _{c}$ and
phase $\phi _{c}$ of the laser. We emphasize that the inseparability of $C_{%
\mathbf{k}\mathbf{q}}$ only implies correlation, and may not give
entanglement depending on the complex phase $\phi $ of $\left\langle \hat{a}%
_{q}(t)\hat{a}_{k}(t)\right\rangle $ through \cite{entanglement
correlation}
\begin{equation}
D=2[1+\bar{n}_{k}+\bar{n}_{q}+2\sqrt{\bar{n}_{k}\bar{n}_{q}(g^{(2)}-1)}\cos
\phi _{kq}  \label{relate}
\end{equation}
which relates the quantity $D$ with normalized correlation $g^{(2)}\doteq
\langle \hat{a}_{k}^{\mathbf{\dagger }}\hat{a}_{q}^{\mathbf{\dagger }}\hat{a}%
_{q}\hat{a}_{k}\rangle /\bar{n}_{q}\bar{n}_{k}$.

\begin{figure}[tbp]
\center\epsfxsize=8.8cm\epsffile{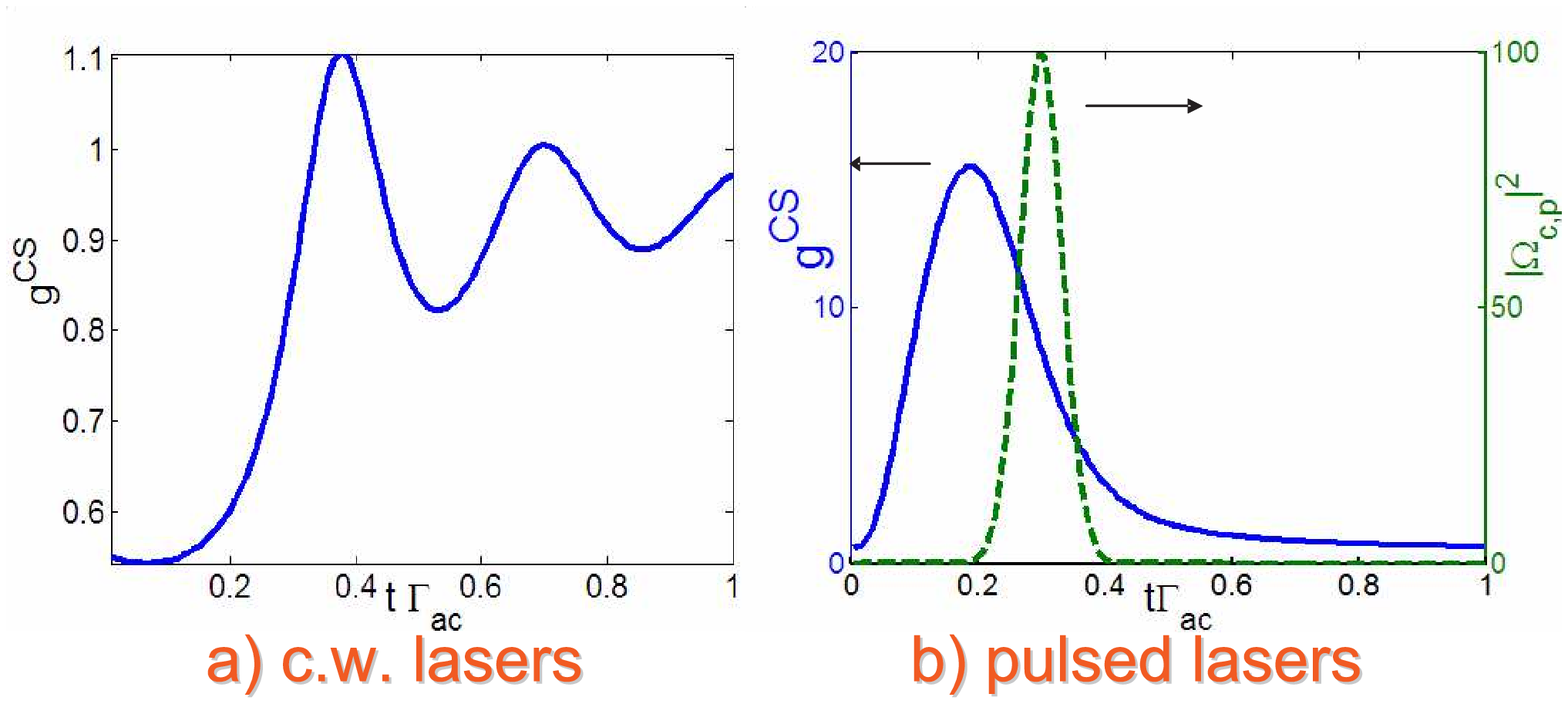}
\caption{(Color online) Effect of pulses on the Cauchy-Schwarz correlation
for DRR ($\Delta _{p}=\Delta _{c}=0,\Omega _{p}=\Omega _{c}$) scheme with:
a) c. w. lasers, b) identical and coincident laser pulses of width $\protect%
\sigma _{c}=1/15\Gamma _{ac}$ (profile shown by the dashed line). Note the
substantial increase in the nonclassical correlation ($g^{CS}>1$) for the
case driven by short pulses. Parameters used are $\Omega _{p,c}=10\Gamma
_{ac}$, no decoherence $\protect\gamma _{bc}=0$ and the initial populations $%
\protect\rho _{cc}(0)=\protect\rho _{bb}(0)=0.5$. We assume Gaussian profile
$\exp [-(t-t_{0})^{2}/\protect\sigma _{c}^{2}]$ for the pulses.}
\label{cwpulseDRR}
\end{figure}


\begin{figure}[tbp]
\center\epsfxsize=8.8cm\epsffile{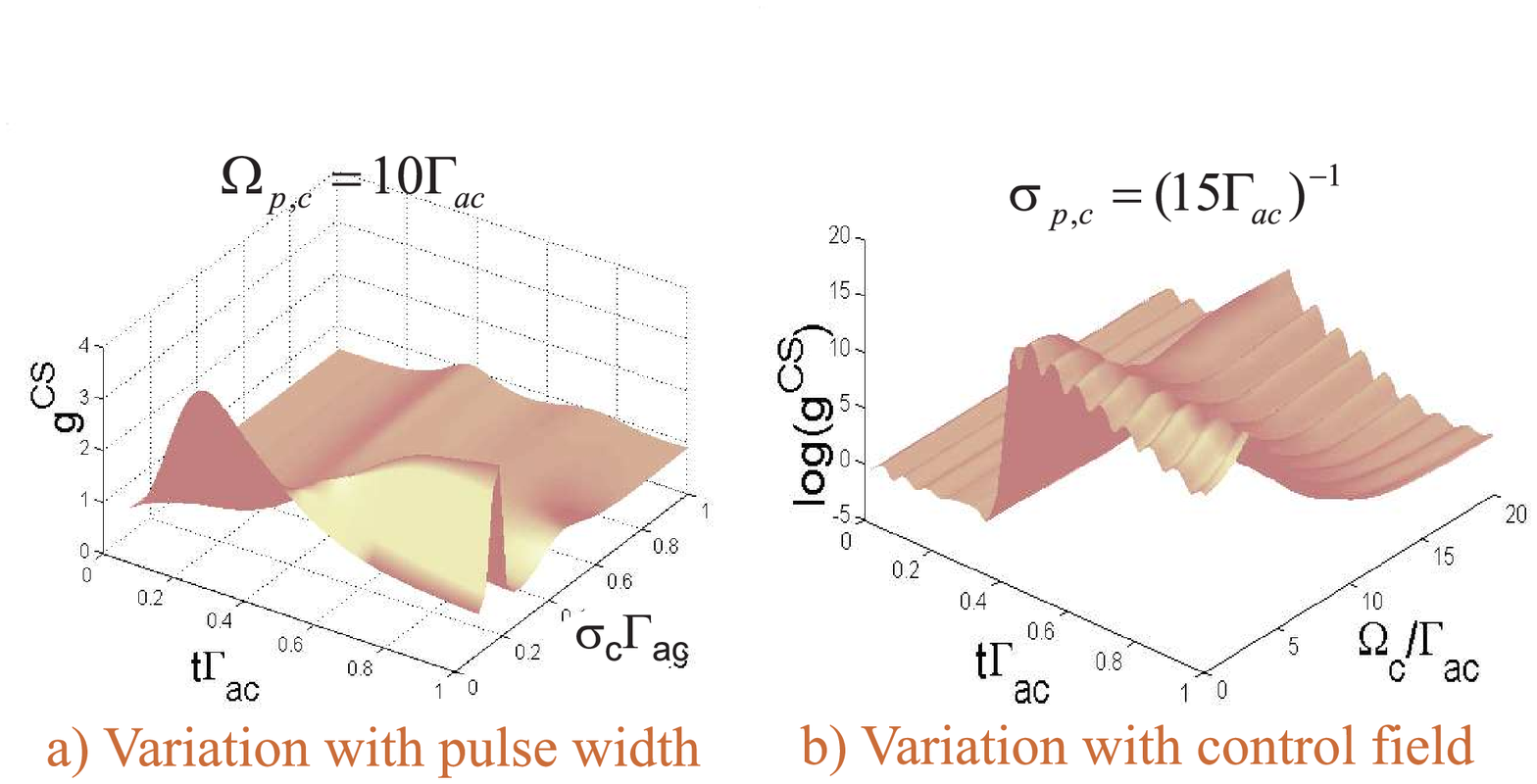} \caption{(Color
online) Variation of the correlation profile with: a) pulse
width $\protect\sigma _{c}$ ($=\protect\sigma _{p}$) for resonant case $%
\Delta _{p,c}=0$, with $\Omega _{p,c}=10\Gamma _{ac}$and b) Rabi frequency $%
\Omega _{c}$ ($=\Omega _{p}$) for detuning $\Delta
_{p,c}=-100\Gamma _{ac}$ and width $\protect\sigma _{c}=1/15\Gamma
_{ac}$. Other parameters are the same as in Fig.
\ref{cwpulseDRR}.} \label{widthRc}
\end{figure}


\section{Results}

Based on the results, we discuss the effects of laser pulses on nonclassical
photon correlation, entanglement and the contribution of quantum noise.

\subsection{Nonclassical photon correlation}

Several interesting effects are found for excitations with ultrashort
pulses. First, let us compare the results (Fig. \ref{cwpulseDRR}) with c.w.
lasers and pulsed lasers for identical and resonant lasers: $\Delta
_{p}=\Delta _{c}=0,\Omega _{p}=\Omega _{c}$ referred to as \emph{double
resonant Raman} (DRR). Continuous wave lasers give a Fresnel-like
oscillations between classical and quantum regimes, showing no apparent
nonclassical correlation (Fig. \ref{cwpulseDRR}a). The oscillations have a
period $\pi /\Omega _{p,c}$, are due to optical nutation \cite{nutation}
which do not show up in the case of pulsed excitations due to the
adiabaticity.

\subsubsection{Effects of finite pulse width}

Laser pulses give nonclassical correlation much larger (Fig. \ref{cwpulseDRR}%
b) than the correlation from c.w. laser. This is one of the main
results. The correlation increases substantially for pulse width
$\sigma _{c}$ shorter than $0.1/\Gamma _{ac}$ (Fig.
\ref{widthRc}a). The physical explanation comes from two points:
a) A short pulse carries a large bandwidth. b) Correlation is a
measure of the amount of spectral interrelation between the Stokes
and anti-Stokes photons.

\subsubsection{Effects of pulse detuning}

Pulses with finite detunings $\Delta _{p,c}>>\Omega _{p,c}$
produce large nonclassical correlation (Fig. \ref{widthRc}b).
Here, the bandwidth of the generated pulses of photons are
enhanced by another mechanism (in addition to the short pulse
duration); i.e. the spontaneous Raman processes from the two laser
pulses. However, for c.w. case, the correlation does not depend on
the Rabi frequencies regardless of whether there is detuning or
not.

\begin{figure}[tbp]
\center\epsfxsize=8.6cm\epsffile{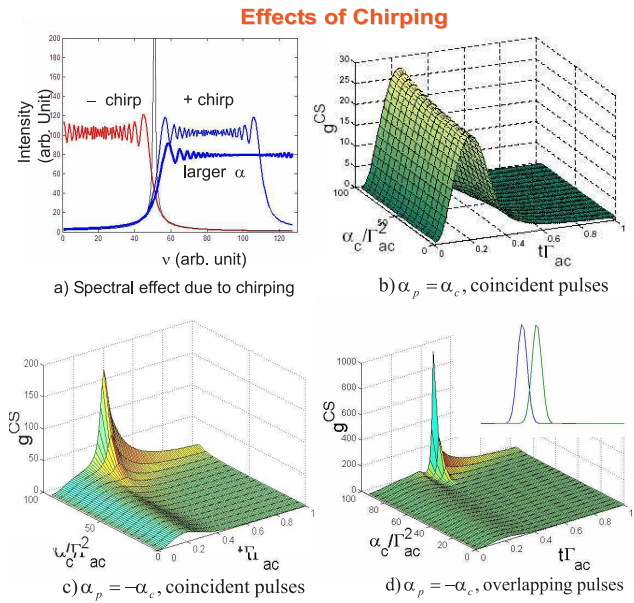} \caption{(Color
online) a) Spectral content of chirped carrier frequency,
computed from the Fourier transform of $\exp [i(\protect\nu _{c}t+\protect%
\alpha _{c}t^{2})]$. Effects of chirping on correlation profile for: b)
coincident pulses and identical chirp, c) coincident pulses and opposite
chirp, d) partially overlapping pulses and opposite chirp. Other parameters
are the same as in Fig. \ref{cwpulseDRR}b.}
\label{chirp}
\end{figure}


\subsubsection{Effects of chirping and delay of pulses}

We now analyze the effects of chirping of the pulses. We consider the
chirped carrier frequency of the form $\nu _{x}(t)=\nu _{x}+\alpha _{x}t$ ($%
x=p,c$). Figure \ref{chirp}a serves to remind the spectral content of
positively chirped and negatively chirped pulses. When the pulses are
chirped identically ($\alpha _{p}=\alpha _{c}$), there is no significant
effect on the correlation (Fig. \ref{chirp}b). For opposite chirping ($%
\alpha _{p}=-\alpha _{c}$) and sufficiently large $|\alpha _{p,c}|$, the
correlation shows a peak. This can be understood qualitatively from Fig. \ref
{chirp}a. The two-spectral ''wings'' span an effectively broad spectrum of
photons that resulted in a large correlation. The correlation can be further
enhanced by displacing the control pulse slightly such that both pulses
overlap partially. This yield an extremely large transient ``correlation
pulse''. If the pulses are entirely separated, the chirping has no
enhancement effect on the correlation.

The Raman-EIT scheme has a very large $g^{CS}$ at times much smaller than
the lifetime $\Gamma ^{-1}$, but decays rapidly to zero for large times,
corresponding to steady state photon antibunching found in previous studies
\cite{HH}.

\subsection{Entanglement}

\begin{figure}[tbp]
\center\epsfxsize=8.0cm\epsffile{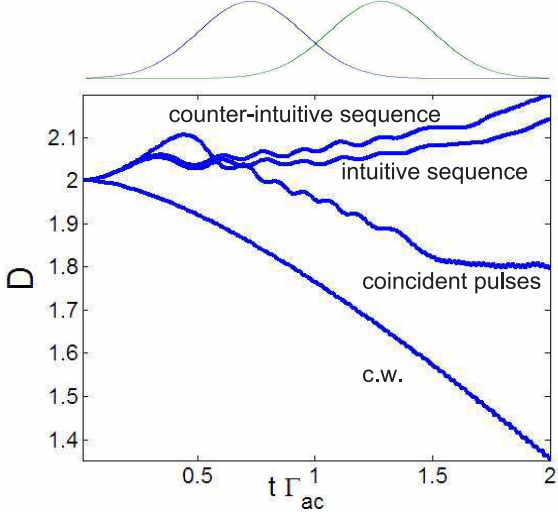}
\caption{(Color online) Entanglement between the Stokes and anti-Stokes
photon pairs for the DRR scheme using: a) continuous wave (c.w.) lasers, b)
coincident pulses, overlapping pulses with c) intuitive sequence, d)
counter-intuitive sequence. The overlapping pulses are shown on top of the
figure. Parameters are: $\protect\sigma _{p,c}=(2.1\Gamma _{ac})^{-1},\Delta
_{p,c}=0,\Omega _{p,c}=30\Gamma _{ac}$, $\protect\gamma _{bc}=0$ and $%
\protect\rho _{cc}(0)=1$. }
\label{entanglepulse}
\end{figure}

Figure \ref{entanglepulse} shows the entanglement for different pulse
sequences for DRR scheme.

\subsubsection{Favorable initial condition}

The \emph{symmetrical} initial condition $\rho _{cc}(0)=\rho _{bb}(0)=0.5$
that gives quantum correlation does not give entanglement. We obtain
entanglement with a different initial condition $\rho _{cc}(0)=1$. This
initial condition ensures that Stokes photon is produced first before an
anti-Stokes photon since there is little or no population in level $b$. But
a more important point that explains entanglement is that, the state of the
atom which generates the anti-Stokes photon carries quantum information
about the Stokes photon. The simultaneous excitations by lasers ensures that
both Stokes and anti-Stokes are driven with the same phase. This explains
how \emph{identical }lasers configuration and $\rho _{cc}(0)=1$ can generate
entangled photon pairs.


\begin{figure}[tbp]
\center\epsfxsize=8.8cm\epsffile{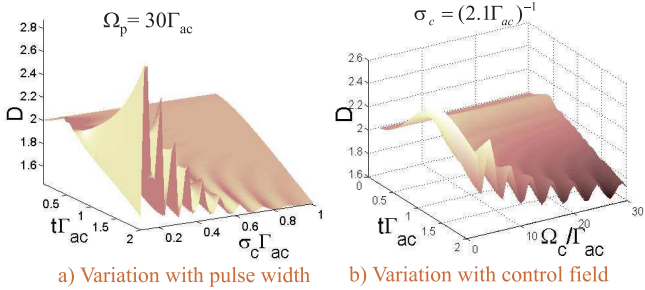} \caption{(Color
online) Variation of the entanglement parameter D for DRR scheme
with: a) pulse width $\protect\sigma _{p,c}$ with $\Omega
_{p,c}=30\Gamma _{ac}$ and b) Rabi frequency $\Omega _{pc}$ with
pulse width $\protect\sigma _{c}=\protect\sigma _{p,c}=(2.1\Gamma
_{ac})^{-1}$ for coincident pulses centered at $t=0.5/\Gamma
_{ac}$. Other parameters are the same as in Fig.
\ref{cwpulseDRR}.} \label{DwidthRc}
\end{figure}


\subsubsection{Effect of pulses on entanglement}

The increased spectral bandwidth due to the pulses may be favorable for the
correlation (see Fig. \ref{cwpulseDRR}b) but not for the entanglement. \emph{%
Coincident pulses }give entanglement while partially overlapping pulses in
both intuitive and counter-intuitive sequences do not give entanglement.

Figure \ref{DwidthRc} illustrates how the entanglement parameter
$D$ varies with pulse width $\sigma _{p,c}$ and laser field
$\Omega _{p,c}$ for DRR scheme. The oscillatory feature implies
that the presence of entanglement could be very sensitive to the
variations of $\sigma _{p,c}$ and $\Omega _{p,c}$. Our analysis
shows that the oscillations in the transient entanglement can be
understood via the pulse area concept. The parameter $D$ is phase
sensitive \cite{entanglement correlation} and would be affected by
the amplitude $\sigma _{p,c}\Omega _{p,c}$ of the pulse area which
manifests as a phase.

\begin{figure}[tbp]
\center\epsfxsize=8.7cm\epsffile{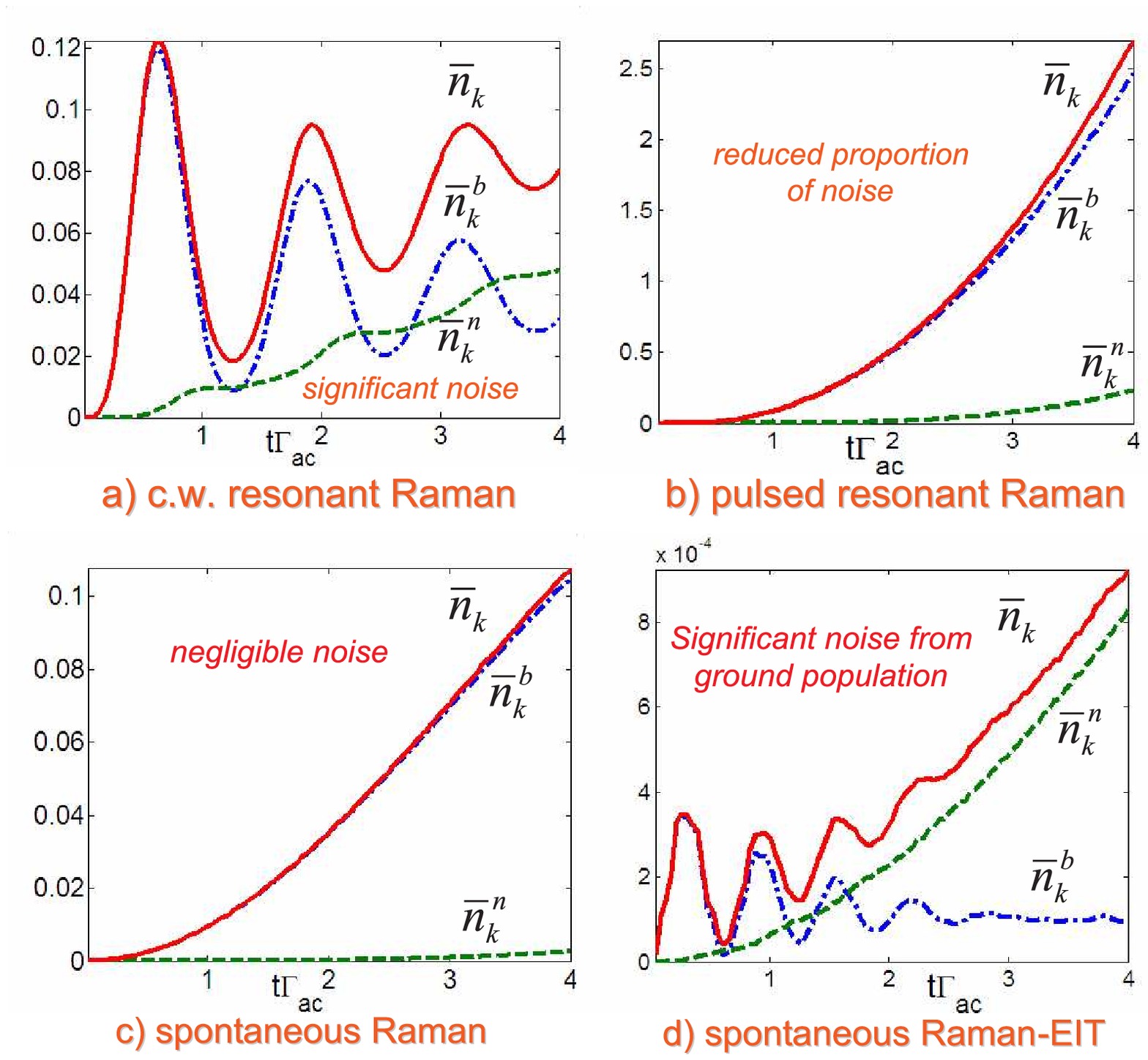}
\caption{(Color online) Mean number of Stokes photon computed from the
solutions with boundary (initial) operators only ($\bar{n}_{k}^{b}$), noise
operators only ($\bar{n}_{k}^{n}$) and both boundary and noise operators ($%
\bar{n}_{k}$) for: a) resonant ($\Delta _{p}=0$) Raman with c.w. pump ($%
\Omega _{p}=5\Gamma _{ac}$), b) resonant Raman with pulsed pump, c)
spontaneous Raman ($\Delta _{p}=-50\Gamma _{ac}$, $\Omega _{p}=5\Gamma _{ac}$%
), b) spontaneous Raman-EIT (add a resonant control laser, $\Omega
_{c}=10\Gamma _{ac}$). We assume $\protect\gamma _{bc}=0$ and the initial
population $\protect\rho _{cc}(0)=1$. }
\label{noiseRaman}
\end{figure}


\subsection{Quantum noise}

Figures \ref{noiseRaman}a,c show the Stokes photon number for single Raman
transition with only the pump field. As expected, the contribution of
quantum noise is significant for resonant pump field (Fig. \ref{noiseRaman}%
a) but negligible for spontaneous Raman (weak and far detuned pump) (Fig.
\ref{noiseRaman}c). This shows that quantum noise is important and
contributes to the photon numbers when the field is resonant and gives rise
to spontaneous emission.

If a pump pulse is used instead, the Stokes photon number increases
significantly but the proportion of the noise contribution is reduced (Fig.
\ref{noiseRaman}b). This can be understood: the pump pulse corresponds to a
range of frequencies, but only the resonant frequency contributes
significantly to the noise through spontaneous emission while most
frequencies that are off-resonant undergo spontaneous Raman process which
gives very little noise.

When a resonant c.w. control field is applied (in addition to the pump
field) we have anti-Stokes photons in the Raman-EIT (electromagnetic induced
transparency) scheme ($\Delta _{p}>>\Omega _{p}$, $\Delta _{c}=0$) \cite
{paper I}. The contribution of the noise to the Stokes photon number becomes
significant despite the large pump detuning (see Fig. \ref{noiseRaman}d).
This seems surprising, but can be explained as due to the \emph{noise in
population} level $c$ due to the spontaneous emission noise of the
anti-Stokes photons.

\section{Conclusions}

To summarize, we have presented a method to obtain exact solutions of photon
operators that include quantum noise. The method enables exact computation
of a variety of quantities that involve products of photon operators such as
photon correlation, entanglement and squeezing in the \emph{transient regime.%
} The results enable us to study the effects of pulsed excitations on the
nonclassical photon correlation and entanglement. Short laser pulses produce
much larger quantum correlation compared to c.w. lasers. The correlation
profile is essentially independent of laser field for resonant pulses.
Finite detuning of the pulses can very large correlation. Most importantly,
we have studied how the laser pulses parameters like chirping, duration and
pulse sequence can give an extremely large nonclassical photon correlation.
Chirping of the pump and the control lasers in an opposite manner increases
the correlation. The correlation increases further for partially overlapping
pulses. We provided explanation based on the spectral content and large
bandwidth of the laser pulses that conspire to enhance the photon
correlation. We learned that the initial condition of the atom determines
the transient entanglement and correlation. We explained how identical\emph{%
\ }lasers configuration and initial condition $\rho _{cc}(0)=1$ can generate
entangled photon pairs. We apply the pulse area concept to explain why the
transient entanglement varies (oscillates) with pulse width and the laser
field. We have also found that the laser pulses tend to reduce the
contribution of quantum noise to the photon number. All these results
provide conceptual insights that show the significance of pulsed excitations
on quantum properties.

\end{document}